 \documentclass[final,5p,times,twocolumn,round]{elsarticle}

\usepackage{natbib}
\usepackage{amssymb}
\usepackage{color}
\usepackage{diagbox}
\usepackage{amsmath}
\usepackage{multirow}
\usepackage{makecell}
\usepackage{slashbox}

\usepackage{caption}

\journal{arXiv}

\begin{document}

\begin{frontmatter}



\title{Adaptive inversion algorithm for 1.5 $\mu m$ visibility lidar incorporating in situ Angstrom wavelength exponent}


\author[label1,label2]{Xiang Shang} 

\author[label1,label2,label3,label4,label5]{Haiyun Xia\corref{cor1}} 
\cortext[cor1]{Corresponding author}
\ead{hsia@ustc.edu.cn}

\author[label1,label2]{Xiankang Dou}

\author[label1,label2]{Mingjia Shangguan}

\author[label1,label2]{Manyi Li}

\author[label1,label2]{Chong Wang}

\author[label1,label2]{Jiawei Qiu}

\author[label1,label2]{Lijie Zhao}

\author[label1,label2]{Shengfu Lin}

\address[label1]{CAS Key Laboratory of Geospace Environment, USTC, Hefei, 230026, China}
\address[label2]{School of Earth and Space Sciences, USTC, Hefei, 230026, China}
\address[label3]{Collaborative Innovation Center of Astronautical Science and Technology, HIT, Harbin, 150001, China }
\address[label4]{Synergetic Innovation Center of Quantum Information and Quantum Physics, USTC, Hefei 230026, China}
\address[label5]{CAS Center for Excellence in Quantum Information and Quantum Physics, USTC, Hefei 230026, China}

\setcitestyle{authoryear}

\begin{abstract}
As one of the most popular applications of lidar systems, the atmospheric visibility is defined to be inversely proportional to the atmospheric extinction coefficient at 0.55 $\mu m$. Since the laser at 1.5 $\mu m$ shows the highest maximum permissible exposure in the wavelength ranging from 0.3 $\mu m$ to 10 $\mu m$, the eye-safe 1.5 $\mu m$ lidar can be deployed in urban areas. In such a case, the measured extinction coefficient at 1.5 $\mu m$ should be converted to that at 0.55 $\mu m$ for visibility retrieval. Although several models have been established since 1962, the accurate wavelength conversion remains a challenge. An adaptive inversion algorithm for 1.5 $\mu m$ visibility lidar is proposed and demonstrated by using the in situ Angstrom wavelength exponent, which is derived from either a sun photometer or an aerosol spectrometer. The impact of the particle size distribution of atmospheric aerosols and the Rayleigh backscattering of atmospheric molecules are taken into account. In comparison, the Angstrom wavelength exponent derived from the sun photometer is 7.7\% higher than that derived from the aerosol spectrometer. Then, using 1.5 $\mu m$ visibility lidar, the visibility with a temporal resolution of 5 min is detected over 48 hours in Hefei (31.83 ${}^ \circ {\rm{N}}$, 117.25 ${}^ \circ {\rm{E}}$). The average visibility error between the new method and a visibility sensor (Vaisala, PWD52) is 5.2\% with the R-square value of 0.96, while the relative error between another reference visibility lidar at 532 nm and the visibility sensor is 6.7\% with the R-square value of 0.91. All results agree with each other well, demonstrating the accuracy and stability of the algorithm.
\end{abstract}

\begin{keyword}


Visibility lidar \sep Aerosol optical depth \sep Aerosol size distribution \sep Angstrom wavelength exponent
\end{keyword}

\end{frontmatter}



\section{Introduction\\ }
\label{}

Visibility, or meteorological optical range (MOR), is of decisive importance for all kinds of traffic operations and air pollution monitoring (Delucchi, Murphy, \& McCubbin, 2002; Werner, Streicher, Leike, \& Munkel, 2005). In the free-space optical (FSO) communication systems, visibility can be used to estimate its availability performance (Yin et al., 2017). According to Koschmieder's theory, MOR is directly related to the extinction coefficient at 0.55 $\mu m$ and to the contrast threshold of an observer who needs to distinguish an object from its background (Baumer et al., 2008).

Lidar measures atmospheric backscattering to retrieve the visibility. It makes observations of atmospheric conditions over an extended optical path from one location, in any, not just in horizontal direction (Weitkamp et al., 2006). Since lidar allows quantitative determination of visual range as a function of distance at any direction, it is ideally suited to monitor visibility at airports.

Lidar systems can make atmospheric observations at different wavelengths ranging from UV to NIR (such as 355 nm, 532 nm and 1064 nm) (Wu et al., 2015; McKendry et al., 2009; Xia et al., 2007). Recently, the 1.5 $\mu m$ lidar has been recognized as an important instrument in the measurements of PM10 (Lisenko, Kugeiko, \& Khomich, 2016), since it studies the Mie backscattering at longer wavelength than those systems mentioned above. By using the stimulated Raman scattering in methane, a transmitter that produces a high-pulse-energy laser at 1.5 $\mu m$ is adopted to develop a direct analog-detection lidar, which has been used for the observation of plumes from aerosol generators (Mayor, Spuler, Morley, \& Loew, 2007; De Wekker, \& Mayor, 2009). Recently, a micro-pulse 1.5 $\mu m$ aerosol lidar is demonstrated to measure the atmospheric parameters in Hefei, China (Xia et al., 2015). 

There are some advantages to monitoring visibility at 1.5 $\mu m$ than at UV and visible wavelengths, including the highest maximum permissible exposure to human eyes (American National Standards Institute, 2007), lower disturbance from Rayleigh backscattering, weaker sky radiance and lower atmospheric attenuation (Liao et al., 2017). Furthermore, 1.5 $\mu m$ is a standard wavelength of optical telecommunications, optical fiber components and devices are commercial available (Shangguan et al., 2016), reducing the cost in constructing a lidar system substantially. Finally, thanks to the lowest attenuation in the optical fiber at 1.5 $\mu m$, an all-fiber integrated system offers unparalleled features in field experiments, such as mechanical decoupling and remote installation of the subsystems, simplification of optical configuration and alignment, and enhancement in coupling efficiency and long-term stability (Xia et al., 2016a).

But, the MOR is commonly defined at 0.55 $\mu m$, the most sensitive wavelength for human eyes. So, all the measured atmospheric extinction coefficients at other wavelengths should be converted to the extinction coefficient at 0.55 $\mu m$. From 1960s, great efforts have been devoted to this issue. A semi-empirical three-stage formula, called the Kruse formula, is typically used to determine the wavelength dependence of the atmospheric extinction coefficient, and it is the only model providing a wavelength dependent relation between the atmospheric visibility and the extinction coefficient during a long time (Kruse, McGlauchlin, \& McQuistan, 1962). At the beginning of the 21st century, Kim indicated that the Kruse formula didn't have a good performance in fog conditions. On the basis of the Kruse formula, Kim gave a five-stage function (Kim, McArthur, \& Korevaar, 2001). Later, Naboulsi considered two specific weather conditions, advection fog and convection fog, with experimental data measured on the site "La Turbie" at Nice, France (Al Naboulsi, 2004). By taking the radius of scatterers into account, Grabner analyzed the wavelength dependence using Mie scattering theory and proposed a new model (Grabner, \& Kvicera, 2011). However, the wavelength dependences are different in these models, because they are derived from experiment data in different continents and different seasons. The aerosol's microphysical properties vary significantly and fast in time and space, making the determination of the wavelength dependence to be a complex problem (Grob et al., 2013). 

A convenient way to convert extinction coefficients between different wavelengths is using Angstrom exponent derived from aerosol optical depth (He et al., 2016; Liang, Zhong, \& Fang, 2006; Doxaran et al., 2002), which is defined as the integration of extinction coefficient along the entire atmospheric column vertically. The aerosol optical depth can be obtained from measurement of sun photometer or satellite remote sensing. A high-precision multiband sun photometer measures the optical properties of the atmosphere based on the sun irradiance and the sky radiance. It provides the quantification and physical-optical characterizations of the aerosols. In an atmospheric correction model, with assumed vertical structure of the atmosphere, aerosol optical depth shown a relation with the meteorological visibility measured horizontally at the surface (Steven, 1998).

In this work, an adaptive inversion algorithm for a 1.5 $\mu m$ visibility lidar is proposed. The Angstrom wavelength exponent is retrieved either from either the aerosol optical depth detected by a sun photometer (Cimel, CE318), or from the particle size distribution (PSD) measured by a particle spectrometer (Grimm, 11-R). Theoretical derivation and experimental comparison between these two methods are conducted. In the experiment, in order to test the accuracy of the algorithm, we compared the extinction coefficients measured by a reference lidar at 532 nm and the 1.5 $\mu m$ lidar. Finally, the visibility measured by 1.5 $\mu m$ lidar is compared with a commercial available visibility sensor (Vaisala, PWD50), demonstrating the correctness and accuracy of the new algorithm.  

\section{Methodology\\}

The lidar equation is given as:

\begin{equation}
N(R) = E{\eta _0}\frac{{{\eta _q}}}{{h\nu }}\frac{A}{{{R^2}}}O(R)\frac{{c\Delta t}}{2}\beta (R)\exp \left[ { - 2\int_0^r {\sigma (R')dR'} } \right],\label{eq:mynum}
\end{equation}
where $N(R)$ is the number of photons backscattered from the range $R$, $E$ is the energy of the laser pulse, ${\eta _0}$ accounts for the optical efficiency of the system, ${\eta _q}$ is the quantum efficiency of the detector, $h$ is the Planck constant, $\nu$ is the frequency of the photon, $A$ is the receiver area of the telescope, $R$ is the range from the lidar to the scattering volume, $O(R)$ is the laser-beam receiver-field-of-view overlap function, $c$ is the speed of light, $\Delta t$ is the duration of the laser pulse, $\beta $ and $\sigma $ are the atmospheric backscatter coefficient and atmospheric extinction coefficient, respectively.

If the atmosphere is homogeneous and $\beta $ does not change with $R$ , the extinction coefficient can be expressed as:

\begin{equation}
\sigma  =  - \frac{1}{2}\frac{{d\ln [{R^2}N(R)]}}{{dR}}.\label{eq:mynum}
\end{equation} 

When measuring the MOR, the laser is emitted horizontally and the atmosphere is approximately homogeneous. Thus we use Eq. (2) to get the extinction coefficient at 1.5 $\mu m$ in this work. However, if the atmosphere is inhomogeneous, inversion algorithms proposed by Klett and Fernald (Klett, 1981; Fernald, 1984) can be used to retrieve the atmospheric extinction coefficient.

According to Koschmieder's theory, visibility is expressed as (Baumer et al., 2008):

\begin{equation}
V = \frac{1}{\sigma }\ln \frac{1}{K} \approx \frac{3}{{{\sigma _0}}},\label{eq:mynum}
\end{equation}
where the contrast threshold $K$ is chosen as 0.05 and the extinction coefficient $\sigma $ is taken at ${\lambda _0} = 550\;nm$. In the practical application of 1.5 $\mu m$ visibility lidar, the measured atmospheric extinction coefficients at ${\lambda _1} = 1548\;nm$ should be converted to the extinction coefficient at 550 nm to calculate visibility.

\subsection{Adaptive inversion algorithm A\\}

A convenient way to convert extinction coefficients at different wavelengths is using Angstrom exponent derived from aerosol optical depth $\tau $:

\begin{equation}
{\alpha _1} =  - \frac{{\ln ({\tau _0}/{\tau _1})}}{{\ln ({\lambda _0}/{\lambda _1})}}.\label{eq:mynum}
\end{equation}
Indexes 0 and 1 stand for wavelength 550 nm and 1548 nm, respectively. The aerosol optical depth is the integration of the extinction coefficient in the vertical direction along the entire atmospheric column:

\begin{equation}
\tau  = \int_0^H {\sigma (h)dh,} \label{eq:mynum}
\end{equation}
where $H$ is the distance from ground to the top of the atmosphere. If we assumed that $\sigma (h)$ can be normalized as:

\begin{equation}
\sigma (h) = \sigma (0)f(h),\label{eq:mynum}
\end{equation}
where $\sigma (0)$ is the extinction coefficient at ground, $f(h)$ represents the change in the extinction coefficient with height. Then, Eq. (4) can be turned into:

\begin{equation}
{\alpha _1} =  - \frac{{\ln ({\sigma _0}/{\sigma _1})}}{{\ln ({\lambda _0}/{\lambda _1})}}.\label{eq:mynum}
\end{equation}
Since the visibility lidar is placed near ground and the laser is emitted horizontally, the extinction coefficient at ground is assumed to be equal to the extinction coefficient derived from Eq. (2). Finally, Eq. (7) is substituted into Eq. (3):

\begin{equation}
V = \frac{3}{{{\sigma _0}}} = \frac{3}{{{\sigma _1}}}{({\lambda _1}/{\lambda _0})^{ - {\alpha _1}}}.\label{eq:mynum}
\end{equation}

It is worth mentioning that only sun direct radiations are used to calculate the Angstrom wavelength exponent in this method. The temporal resolution of this method is about 10 minutes. On the other hand, from the combination of direct and diffuse solar radiations measured by sun photometer, the columnar PSD can be retrieved (Nakajima et al., 1996). It is based on the definition of aerosol optical depth:

\begin{equation}
{\tau _\lambda } = \int_{{r_{\min }}}^{{r_{\max }}} {\pi {r^2}} {Q_{ext}}\left[ {\frac{{2\pi r}}{\lambda },m_\lambda ^ * } \right]{n^ * }(r)dr,\label{eq:mynum}
\end{equation}
where $r$ is the particle's radius, ${Q_{ext}}$ is the extinction efficiency factor, $m_\lambda ^*$ is the complex refractive index at wavelength $\lambda $, ${n^*}(r)$ is columnar radius distribution of aerosol. As an example, the ${Q_{ext}}$ is calculated using Mie scattering theory when $m_\lambda ^* = 1.33$, as shown in Fig. 1(a) (Matzler, C., 2002). Few different inversion algorithms have been proposed over the years. Nakajima implemented an algorithm called SKYRAD, which was developed to be applied on Prede-POM sun photometers. A pre-processing module was implemented to adapt the Cimel CE318 sun photometer data to the SKYRAD version 4.2 algorithm (Estelles et al., 2012). In this work, we use SKYRAD version 4.2 to get the PSD from measurements of Cimel CE318 sun photometer. A bimodal lognormal distribution is used to fit the PSD raw data:

\begin{equation}
\begin{aligned}
&{n^*}(r) = \frac{{{C_1}}}{{\sqrt {2\pi } {\delta _1}r}}\exp \left[ { - \frac{{{{(\ln r - \ln {R_1})}^2}}}{{2\delta _1^2}}} \right]\\
& + \frac{{{C_2}}}{{\sqrt {2\pi } {\delta _2}r}}\exp \left[ { - \frac{{{{(\ln r - \ln {R_2})}^2}}}{{2\delta _2^2}}} \right],\label{eq:mynum}
\end{aligned}
\end{equation}
where ${C_1}$, ${\delta _1}$, ${R_1}$, ${C_2}$, ${\delta _2}$, ${R_2}$ are the fitting parameters of the bimodal lognormal distribution. The volume size distribution is also widely used in the atmospheric researches. It is defined as:

\begin{equation}
d{V^*}/d\log (r) = \frac{4}{3}\pi {r^3}{n^*}(r)dr/d\log (r).\label{eq:mynum}
\end{equation}

Two typical bimodal lognormal size distributions in Hefei (31.83 ${}^ \circ {\rm{N}}$, 117.25 ${}^ \circ {\rm{E}}$) are shown in Fig. 1(b). If the index of refraction and PSD in Eq. (9) are determined, the aerosol optical depth versus wavelength can be calculated, as shown in Fig. 1(c). Then the Angstrom wavelength exponent ${\alpha _1}$ can be calculated by using Eq. (4). But the temporal resolution of this method is about 30 minutes, because sky diffuse radiations are indispensable for getting columnar radius distribution of aerosol.

\begin{figure}[h]
\centering
\resizebox{8.5cm}{!}{\includegraphics{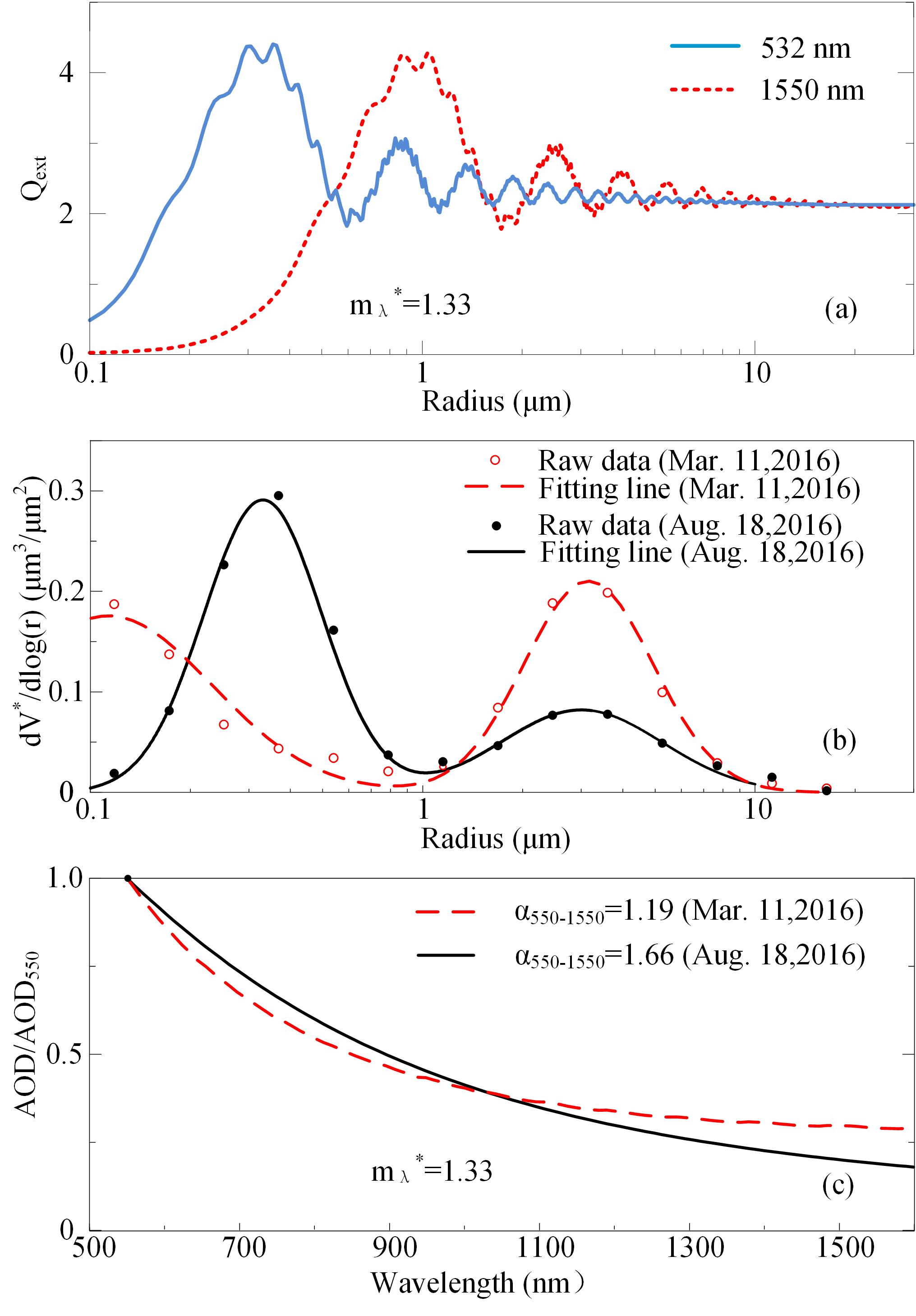}}
\caption{\label{1f}The extinction efficiency factors of 532 nm and 1.55 $\mu m$ (a), two types of typical bimodal lognormal size distributions in Hefei (b), and normalized aerosol optical depth (c).}
\end{figure}

\subsection{Adaptive inversion algorithm B\\}

The extinction coefficient $\sigma _\lambda ^{'}$ due to Mie scattering of aerosol particles can be calculated as:

\begin{equation}
\sigma _\lambda ^{'} = \int_{{r_{\min }}}^{{r_{\max }}} {\pi {r^2}{Q_{ext}}\left[ {r,\lambda ,m_\lambda ^*} \right]n(r)dr} ,\label{eq:mynum}
\end{equation}
where the particle size distribution $n(r)$ is the number of particles per unit volume in the interval $(r,r + dr)$. For a given wavelength $\lambda $, there are only two unknowns $n(r)$ and $m_\lambda ^*$ in Eq. (12). It is worth mentioning that Mie scattering theory applies to spherical particles. For non-spherical particles like dust, other theories should be used (Luo et al., 2015).

When $n(r)$ is measured by aerosol spectrometer and the refractive index is an empirical value, $\sigma _0^{'}$ and $\sigma _1^{'}$ can be calculated using Eq. (12). Then, the Angstrom wavelength exponent can be calculated as:

\begin{equation}
{\alpha _2} =  - \frac{{\ln (\sigma _0^{'}/\sigma _1^{'})}}{{\ln ({\lambda _0}/{\lambda _1})}}.\label{eq:mynum}
\end{equation}

Using the atmospheric extinction coefficient ${\sigma _1}$ obtained by Eq. (2) and the Angstrom wavelength exponent ${\alpha _2}$ obtained by Eq. (13), ${\sigma _0}$ is calculated as:

\begin{equation}
{\sigma _0} = {\sigma _1}{({\lambda _1}/{\lambda _0})^{{\alpha _2}}}.\label{eq:mynum}
\end{equation}

The visibility can be calculated by:

\begin{equation}
V = \frac{3}{{{\sigma _{{\lambda _0}}}}} = \frac{3}{{{\sigma _{{\lambda _1}}}}}{({\lambda _1}/{\lambda _0})^{ - {\alpha _2}}}.\label{eq:mynum}
\end{equation}

In comparison, the sun photometer is a passive detection instrument with lower time resolution than aerosol spectrometer, and it is affected by the weather conditions. Thus, adaptive inversion algorithm B is expected to have a better performance in data processing than algorithm A.

\subsection{Correction of Rayleigh backscattering\\}

It is important to note that the wavelength dependence of molecular Rayleigh scattering is different from aerosol Mie scattering (Xia et al., 2012). The molecular Rayleigh extinction coefficient ${\sigma _{R,\lambda }}$ is approximated as (US Standard Atmosphere, 1976):

\begin{equation}
{\sigma _{R,\lambda }} = \frac{{9.807}}{{{{10}^{20}}}}\frac{{273}}{T}\frac{P}{{1013}}{(\frac{{{{10}^7}}}{\lambda })^{4.0117}},\label{eq:mynum}
\end{equation}
where $T$ and $P$ are the atmospheric temperature and pressure, respectively. As Rayleigh extinction is considered, Eq. (15) should be updated to calculate the atmospheric visibility:

\begin{equation}
V = \frac{3}{{({\sigma _1} - {\sigma _{R,1}}){{({\lambda _1}/{\lambda _0})}^{\alpha _2}} + {\sigma _{R,0}}}}.\label{eq:mynum}
\end{equation}

\section{Instrumentation\\ }

The 1.5 $\mu m$ lidar system used in this experiment has been described in detail elsewhere for aerosol and wind detection (Xia et al., 2015; Xia et al., 2016a). A brief review is given here. The lidar operating at 1.5 $\mu m$ ensures the eye safety of human beings, so that the experiment can be performed in urban areas. As detectors are considered, InGaAs/InP avalanche photodiode is used for 1.5 $\mu m$ detection commonly, but it suffers low efficiency (about 10\%), high noise (a few kHz) and high after pulsing possibility (18\%) (Yu, 2017) .To solve this problem, an up-conversion detector (UCD) is used in this experiment. The UCD up-converts photons at 1.55 $\mu m$ to 863 nm in a periodically poled lithium niobate waveguide (PPLN-W). Then single photon at 1.55 $\mu m$ can be counted by using a Si: APD with high efficiency (20\%), low noise (300 Hz) and negligible after pulsing possibility (0.2\%) (Shentu et al., 2013; Xia et al., 2016b). 

The 532nm lidar is a mature system, which has been used for atmospheric detection for decades. A Mie scattering lidar system at 532 nm is built for observation of optical properties in the atmospheric boundary layer. The key parameters of two lidars are listed in Table 1. The 532 nm lidar has higher pulse energy and larger telescope than the 1.5 $\mu m$ lidar.

\begin{table}[!htbp]
\renewcommand{\arraystretch}{1.3}
\centering
\caption{Key parameters of the visibility lidars.} \label{KeyRate}
\begin{tabular}{p{3.8cm}p{1.9cm}p{1.9cm}}
\hline
\small Parameter & 1.5 $\mu m$ lidar & 532 nm lidar  \\
\hline
\small Wavelength (nm) & 1548 & 532  \\
\small Pulse duration (ns) & 300 & 7  \\
\small Pulse energy (mJ) & 0.11 & 15  \\
\small Pulse repetition rate (kHz) & 15 & 0.05  \\
\small Collimator aperture (mm) & 100 & 80  \\
\small Coupler aperture (mm) & 80 & 300  \\
\small Fiber diameter ($\mu m$) & 10 & 1500  \\
\small Fiber attenuation (dB/km) & 0.02 & 30  \\
\small Detector efficiency (\%) & 20 & 40  \\
\small Dark count noise (Hz) & 300 & 100  \\
\hline
\end{tabular}
\end{table} 

The sun photometer (Cimel, CE318) is a benchmark device for most aerosol observing networks and more specifically for the international federation of AERONET. In this experiment, a CE318N-EDPS9 sun photometer is used for atmospheric observation, which can get the aerosol optical depth at different wavelengths (340 nm, 380 nm, 440 nm, 500 nm, 670 nm, 870 nm, 1020 nm, 1640 nm) and sky radiance data (440 nm, 500 nm, 670 nm, 870 nm, 1020 nm and 1640 nm).

The aerosol spectrometer (Grimm, 11-R) combines optical single particle detection for counting and classifying aerosol particles. It uses light source at a wavelength of 660 nm and measures the scattering light at $90^{\circ}$ . It has 31 size channels with a particle detection size ranging from 0.25 $\mu m$ to 32 $\mu m$. The temporal resolution of Grimm 11-R is 6 s. In this experiment, the raw data is averaged over 5 minutes.

The visibility sensor used in this experiment is a popular present weather detector (Vaisala, PWD50). The visibility sensor combines the functions of a forward scatter visibility meter, and evaluates MOR by measuring the intensity of infrared light scattered at a wavelength of 875 nm at an angle of 45${}^ \circ $. The scatter measurement is converted to the visibility value, with an accuracy of $ \pm $ 10\% at the 10 to 10000 m range and $ \pm $ 20\% at the 10000 to 35000 m range. Fig. 2 is a photo of the instruments used in the field experiments.

\begin{figure}[h]
\centering
\resizebox{8.5cm}{!}{\includegraphics{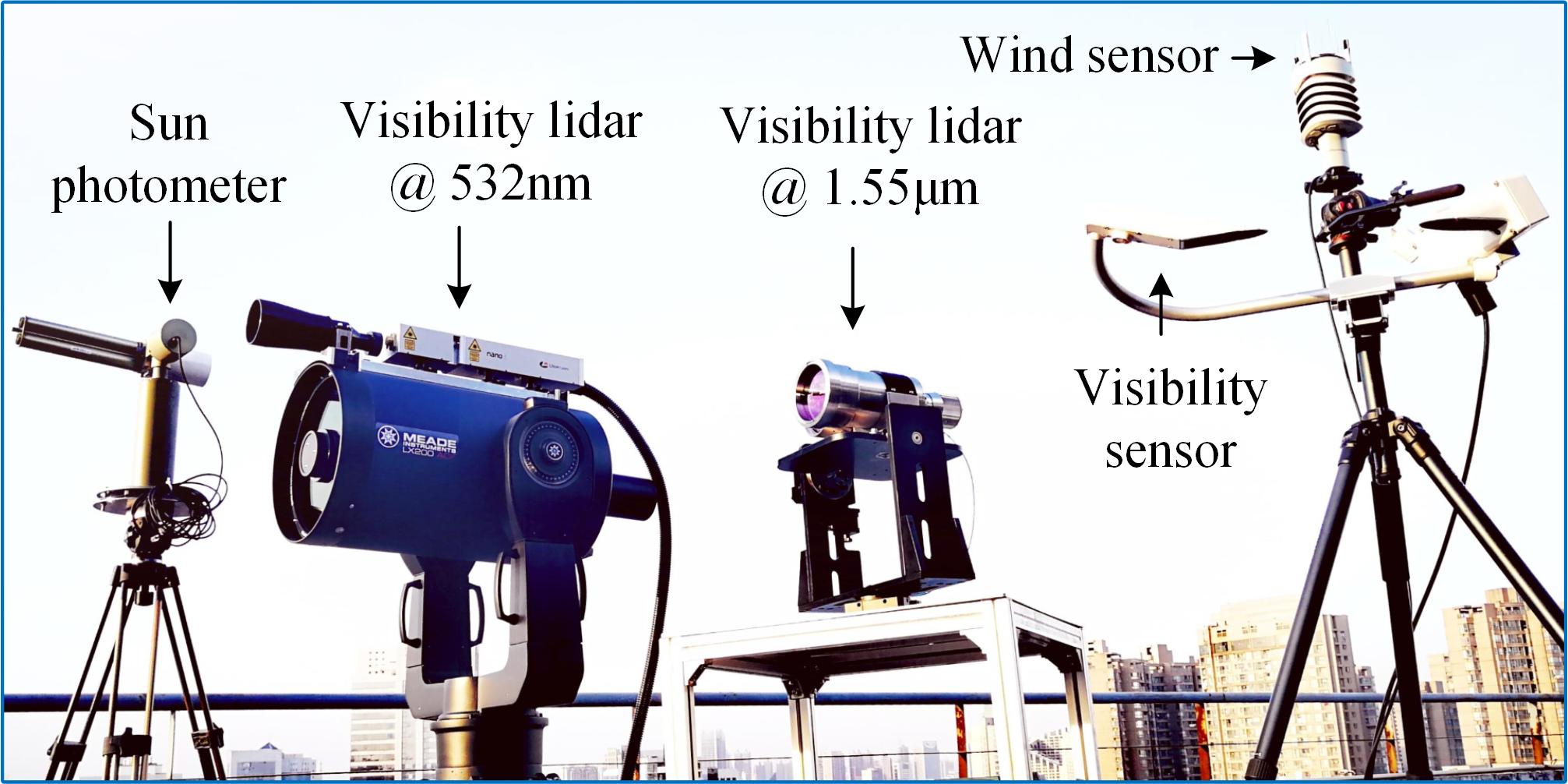}}
\caption{\label{1f}Instruments used in the field experiments.}
\end{figure}

\section{Field experiments and results\\ }

From 9:00 on April 21 to 9:00 on April 23, 2017, atmospheric visibility is detected at Hefei (31.83 ${}^ \circ {\rm{N}}$, 117.25 ${}^ \circ {\rm{E}}$) in Anhui province, China to demonstrate the new proposed algorithm. The location is 40 m above the sea level. The visibility can be calculated by using either Eq. (8) or Eq. (17), and compares with the visibility measured by the Vaisala PWD50.

Raw lidar backscattering signal of 1.5 $\mu m$ visibility lidar over 48 hours is shown in Fig. 3(a). The temporal resolution and spatial resolution are set to 5 min and 45 m, respectively. Although the pulse energy is 110 $\mu J$ and the diameter of the telescope is only 80 mm, it can be seen in Fig. 3(a) that the signal can extend to 9 km horizontally. The pulse energy of the 532 nm lidar is 15 mJ, and the aperture of the telescope is 300 mm. As shown in Fig. 3(b), the raw backscattering signal at 532 nm lidar only extends to about 6 km horizontally. It is mainly due to the fact that, the extinction coefficient at 532 nm is larger than that at 1550 nm for the same atmospheric conditions. Through the decaying speed of raw lidar signal at 532 nm, it is clear that the visibility during the first night changes fast, while it is relatively stable during the second night.

\begin{figure}[!htbp]
\centering
\resizebox{8.8cm}{!}{\includegraphics{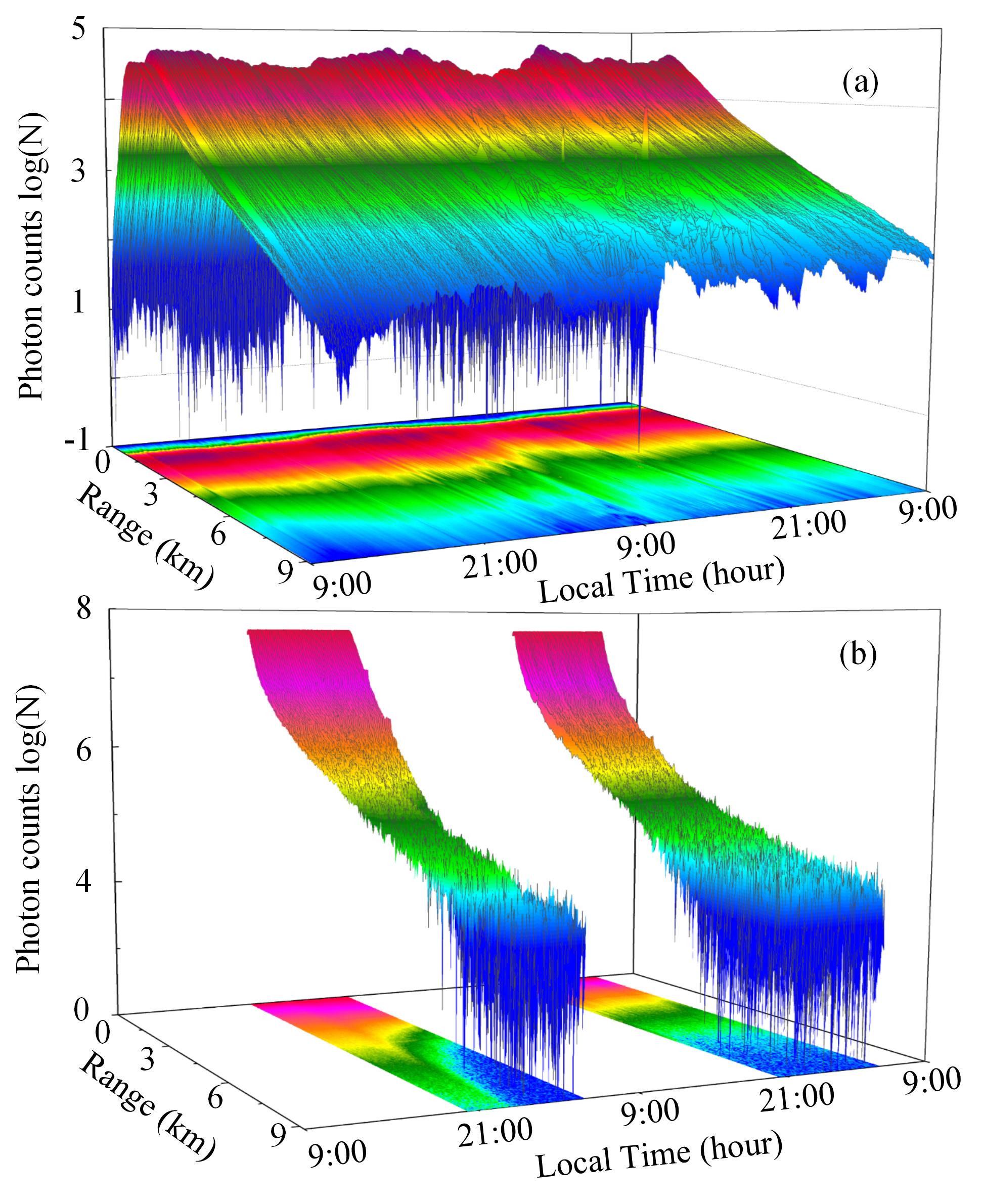}}
\caption{\label{1f}Raw data of 1.55 $\mu m$ lidar (a) and 532 nm lidar (b). }
\end{figure}

The retrieved extinction coefficients at two wavelengths are plotted in Fig. 4. In the two-day experiment, the value of ${\sigma _{1548}}$ varies between 0.033 ${\rm{k}}{{\rm{m}}^{ - 1}}$ and 0.235 ${\rm{k}}{{\rm{m}}^{ - 1}}$. While the values of ${\sigma _{532}}$ varies between 0.132 ${\rm{k}}{{\rm{m}}^{ - 1}}$ and 0.585 ${\rm{k}}{{\rm{m}}^{ - 1}}$, and between 0.203 ${\rm{k}}{{\rm{m}}^{ - 1}}$ and 0.328 ${\rm{k}}{{\rm{m}}^{ - 1}}$ for two nights, respectively. The ${\sigma _{532}}$ is always larger than ${\sigma _{1548}}$ throughout the experiment. A process of declining visibility is observed from 20:15, Apr.21, 2017 to 4:55 next morning. Comparing the extinction coefficients at 20:15 and 4:55, the relative change of ${\sigma _{1548}}$ is 57\%, on the contrary, the relative change of ${\sigma _{532}}$ is 246\%.

\begin{figure}[h]
\centering
\resizebox{8.5cm}{!}{\includegraphics{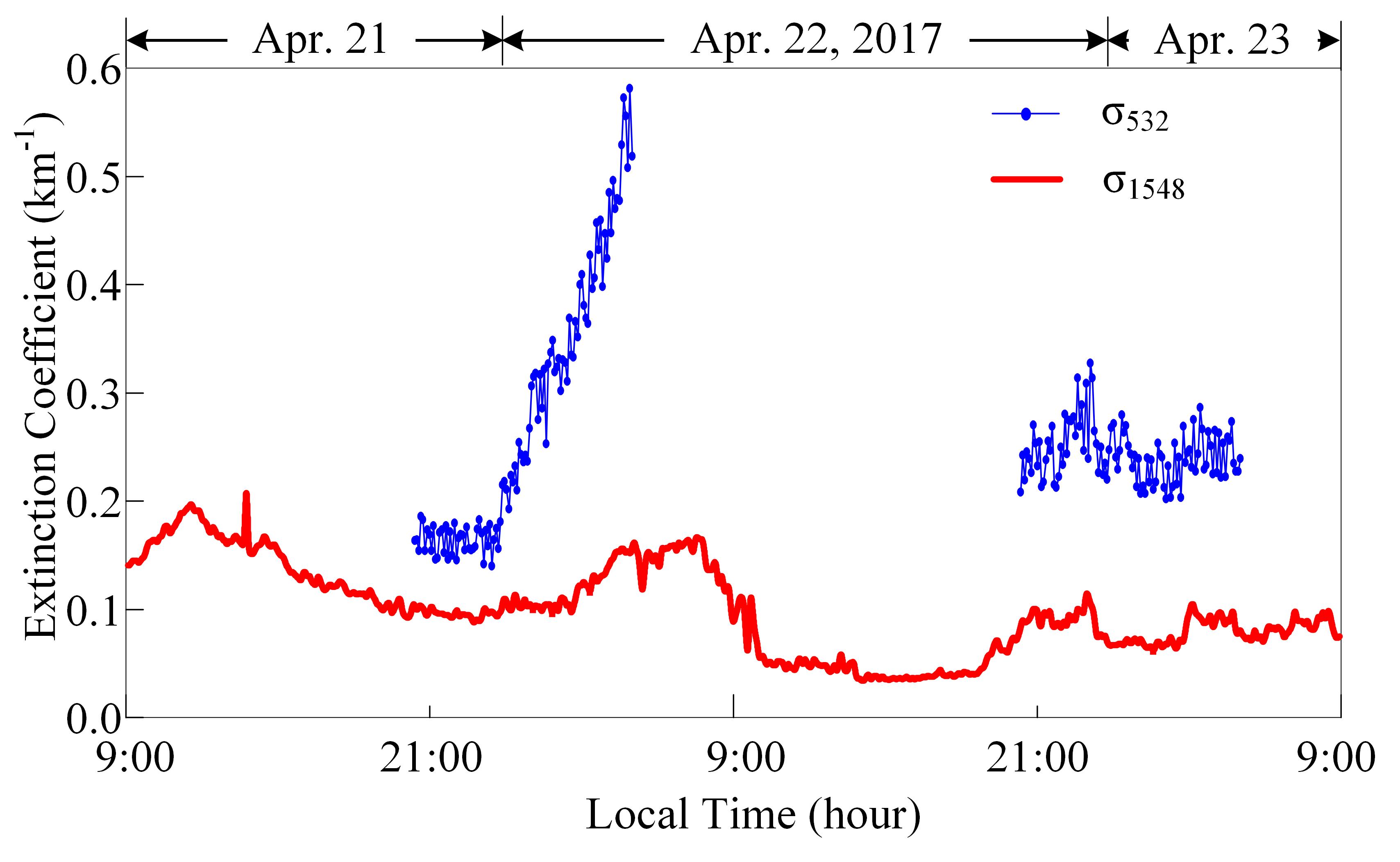}}
\caption{\label{1f}The extinction coefficients of 1.55 $\mu m$ lidar and 532 nm lidar.}
\end{figure}

The PSD is obtained from the aerosol spectrometer. To see changes of PSD more intuitively, we convert number distributions $n(r)$ to volume distributions by using the following equation:

\begin{equation}
dV/d\log (r) = \frac{4}{3}\pi {r^3}n(r)dr/d\log (r).\label{eq:mynum}
\end{equation}

The volume distributions over 48 hours are shown in Fig. 5. The fine mode and coarse mode have different trends. The change period of fine mode's volume distribution is 24 hours and there are two peaks in the 48 hours observation. It keeps lower than 0.03 ${\rm{\mu }}{{\rm{m}}^{\rm{3}}}{\rm{/m}}{{\rm{m}}^{\rm{3}}}$ from 9:00 on Apr.21 to 21:00 on Apr.21. Then it becomes larger over next 4 hours and keeps higher than 0.05 ${\rm{\mu }}{{\rm{m}}^{\rm{3}}}{\rm{/m}}{{\rm{m}}^{\rm{3}}}$ till 9:00 on Apr.21. The next day and the first day have almost the same variation. Generally, the amount of fine mode's particles rises from 21:00, then it keeps a larger value than daytime. However the coarse mode's volume distributions are affected by emergencies and less regular. From 9:00 on Apr.21 to 17:00 on Apr.21, a burst of coarse mode's particles is observed.

\begin{figure}[h]
\centering
\resizebox{8.5cm}{!}{\includegraphics{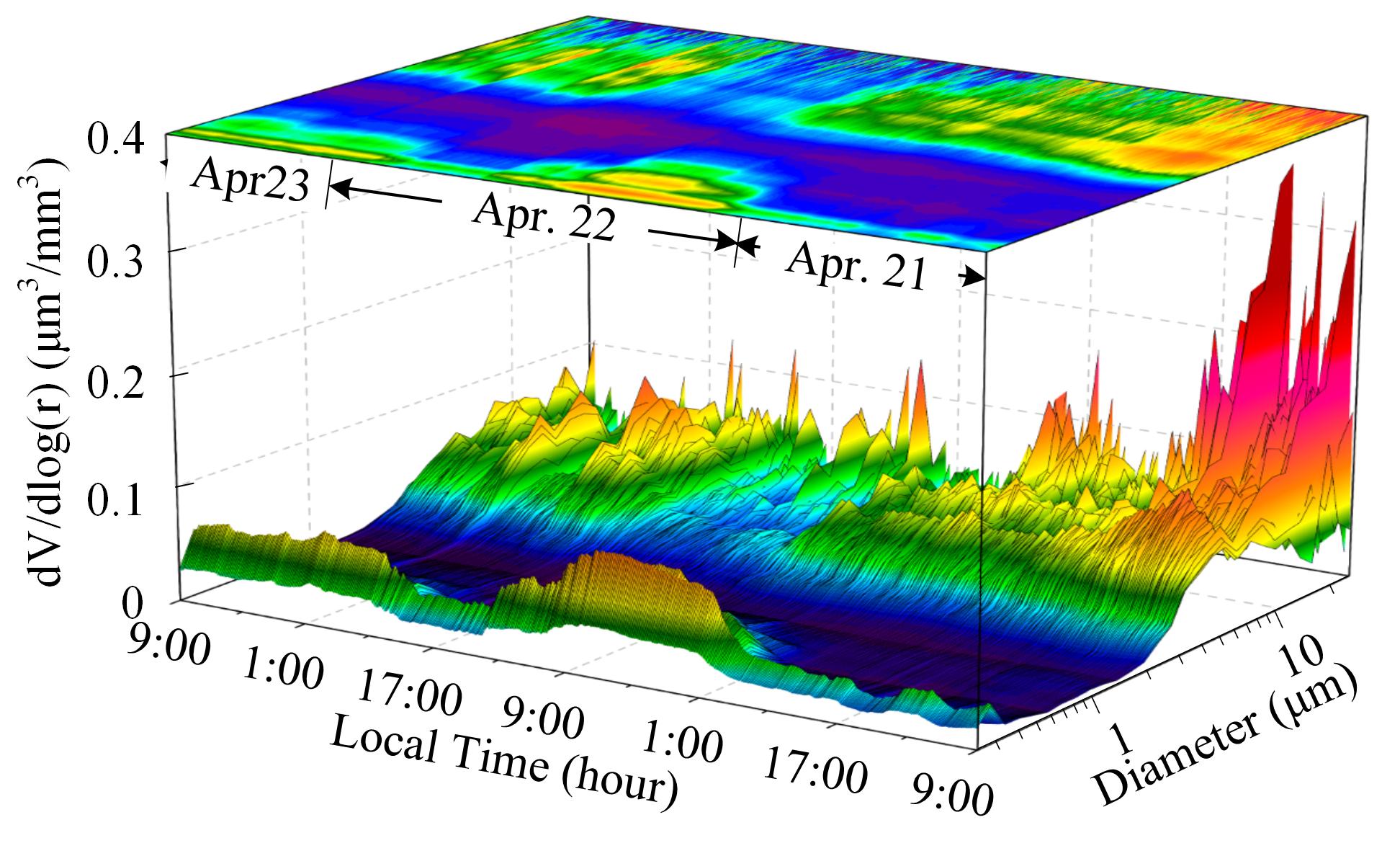}}
\caption{\label{1f}Volume size distribution during 48 hours.}
\end{figure}

The Angstrom wavelength exponent for wavelengths 0.55 $\mu m$ and 1.55 $\mu m$ derived from both aerosol spectrometer and sun photometer are shown in Fig. 6. The Angstrom wavelength exponent from aerosol spectrometer rises from 0.40 at 9:00, Apr.21 to 1.31 at 9:00, Apr.22. The Angstrom wavelength exponent is smaller than 0.5 during the first 12 hours since the air has a lot of coarse mode’s particles. There are 27 sets of Angstrom wavelength exponents derived from sun photometer. Except 5 sets of them are almost equal to the value derived from the aerosol spectrometer, the other 22 sets of them are higher (about 7.7\%) than corresponding values derived from aerosol spectrometer. The reason of this phenomenon is that sun photometer measures the entire atmospheric column, while aerosol spectrometer is an in situ instrument measures the layer above the ground. In general, due to the influence of gravity, the relative content of coarse mode’s particles in high altitude is lower than that on surface, which makes the column-average Angstrom wavelength exponent bigger than Angstrom wavelength exponent measured on surface.

\begin{figure}[h]
\centering
\resizebox{8.5cm}{!}{\includegraphics{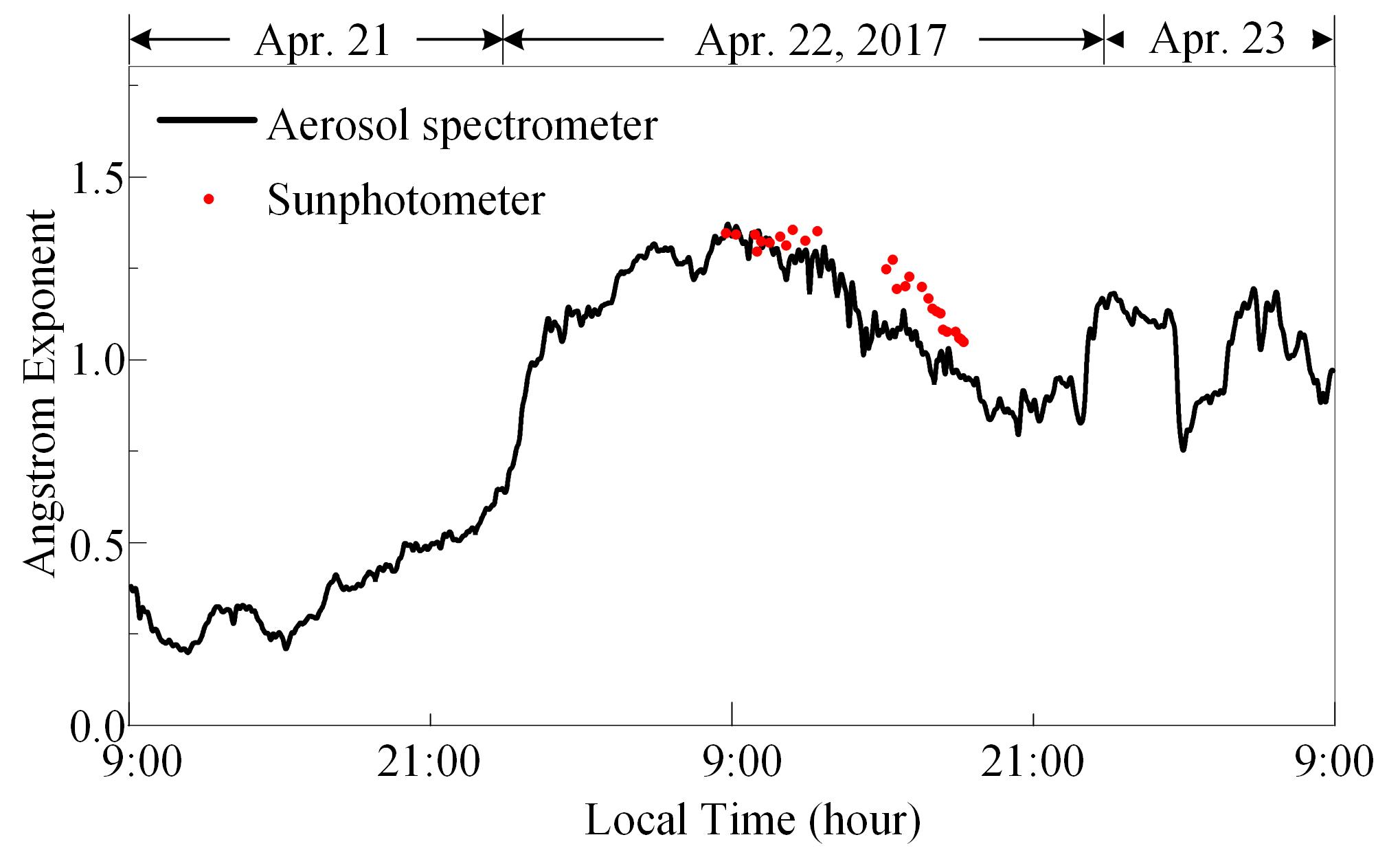}}
\caption{\label{1f}The Angstrom wavelength exponent derived from aerosol spectrometer and sun photometer.}
\end{figure}

The visibilities derived from 1.5 $\mu m$ eye-safe visibility lidar using three different methods and Vaisala PWD50 are shown in Fig. 7(a). The two segments of visibility measured by 532 nm visibility lidar are also shown in Fig. 7(a). 

As we can see from Fig. 7(a), visibility is 14.4 km when the experiment beginning. Then visibility rises to 17.9 km until 23:00, Apr.21. Between 23:00, Apr.21 and 8:00, Apr.22, the visibility falls fast as the emergence of haze. Because of the evaporation of water, visibility rises to 25.9 km after the sunrise. It’s obvious that the visibility of the second day is higher than the visibility at the same time of the previous day, because the particles’ amounts of the second day are smaller than the first day, as shown in Fig. 5. Errors in estimating the visibility at 532 nm during 48 hours are shown in Fig. 7(b). The results are highly consistent between 532 nm lidar and visibility sensor with a relative error of 6.7\%, as well as the R-square value of 0.91.

Errors in estimating the visibility at 1.5 $\mu m$ during 48 hours for four models are shown in Fig. 7(c), (d) and (e). The visibility derived from 1.5 $\mu m$ lidar using Kruse's model and Grabner's model are shown in Fig. 7(c). The R-square is -0.37 and the relative error is 30.2\% for Kruse's model. As for Grabner's model, the R-square is -0.15 and the relative error is 35.0\%. As a comparison, the results of algorithm A is shown in Fig. 7(d). The results of algorithm A have a better performance than Kruse's and Grabner's work. The R-square is 0.87 and the relative error is 7.9\%. As can be seen from Fig. 7(e), the results of algorithm B have the best performance in contrast to the other models. The R-square is 0.96 and the relative error is 5.2\%.The extinction coefficients of 1548 nm used in the four methods are the same. It's obvious that, with the reference of in situ measurements from PWD50, algorithm B is the most accurate algorithm.

\begin{figure*}[!htbp]
\centering
\resizebox{12.0cm}{!}{\includegraphics{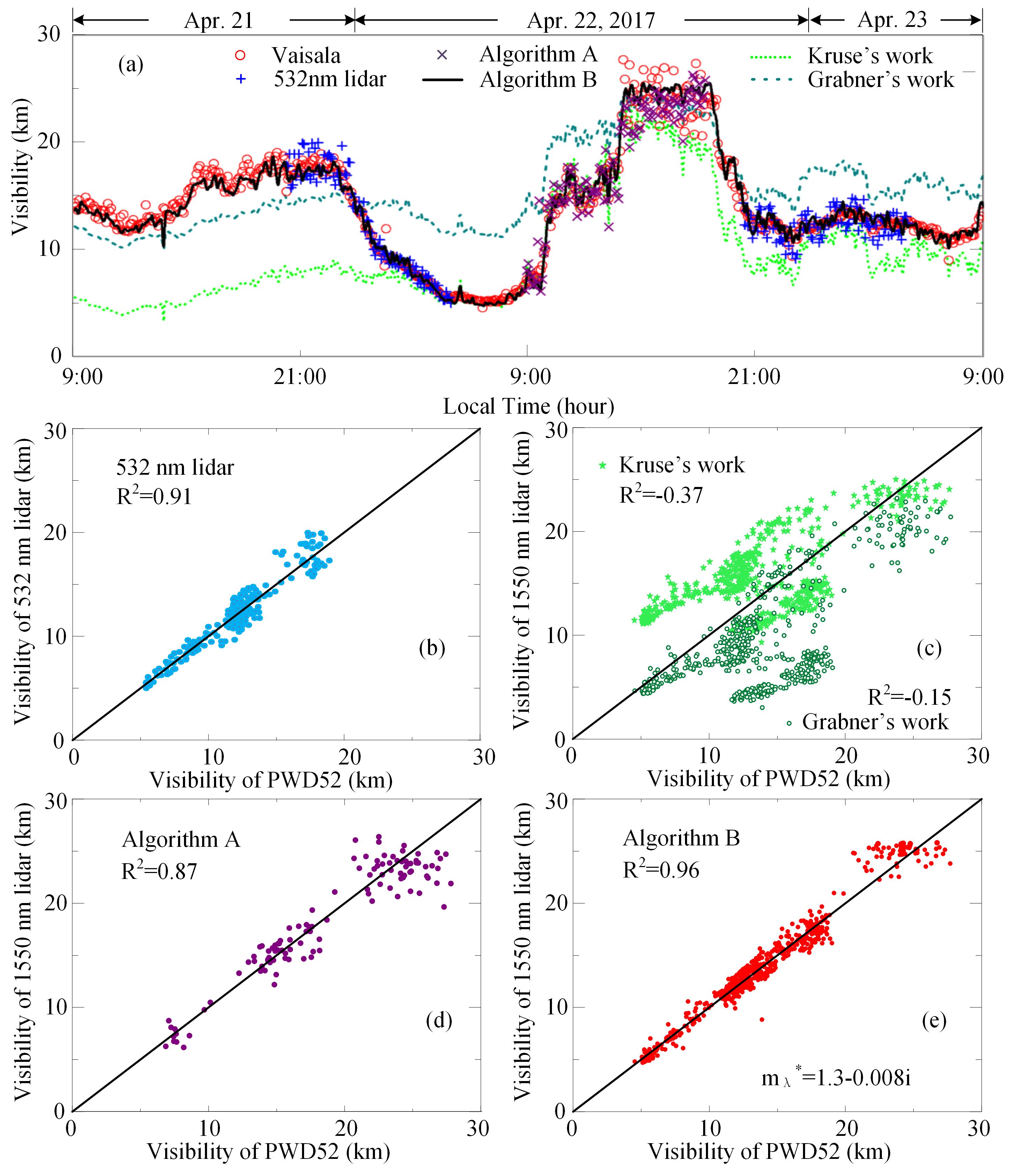}}
\caption{\label{1f}Visibilities derived from different methods (a) , 532 nm lidar’s measurement (b), comparison of visibilities based on Kruse’s work and Grabner’s work (c), Algorithm A (d) and Algorithm B (e).}
\end{figure*}

\section{ Conclusion\\ }

A method for the retrieval of visibility from 1.5 $\mu m$ lidar combined with the measurements of aerosol spectrometer was proposed, which can adjust measures to local conditions. A compact micro-pulse aerosol lidar incorporating a fiber laser at 1.5 $\mu m$ has been constructed in order to verify this method. A 532 nm lidar also has been constructed to compare the atmospheric extinction coefficient between 532 nm and 1.55 $\mu m$. And a Vaisala visibility sensor was used to validate our new model. Continuous observation of visibility was performed. In the comparison experiments, the average relative error between the retrieved visibility using our method and visibility measured by visibility sensor is 5.2\%. If there is no significant pollution sources along the detection path, our method shows good agreement with traditional visibility sensor during day and night. Experiments also demonstrated that the calculation of atmospheric optical properties with the combination of aerosol spectrometer and Mie-scattering theory has high accuracy.

\section*{ Acknowledgments\\ }

We thank Prof. Yunfei Fu, Prof. Renmin Yuan, Prof. Yu Wang and Dr. Hao Liu from the University of Science and Technology of China for their support in experiment and useful discussion.

\section*{ References\\ }
\small
Al Naboulsi, M. (2004). Fog attenuation prediction for optical and infrared waves. Optical Engineering, 43(2), 319-329.

American National Standards Institute (2007). American national standard for the safe use of lasers. ANSIZ136. 

Baumer, D., Vogel, B., Versick, S., Rinke, R., Möhler, O., \& Schnaiter, M. (2008). Relationship of visibility, aerosol optical thickness and aerosol size distribution in an ageing air mass over South-West Germany. Atmospheric Environment, 42(5), 989-998.

Delucchi, M. A., Murphy, J. J., \& McCubbin, D. R. (2002). The health and visibility cost of air pollution: a comparison of estimation methods. Journal of Environmental Management, 64(2), 139-152.

De Wekker, S. F., \& Mayor, S. D. (2009). Observations of atmospheric structure and dynamics in the Owens Valley of California with a ground-based, eye-safe, scanning aerosol lidar. Journal of Applied Meteorology and Climatology, 48(7), 1483-1499.

Doxaran, D., Froidefond, J. M., Lavender, S., \& Castaing, P. (2002). Spectral signature of highly turbid waters: Application with SPOT data to quantify suspended particulate matter concentrations. Remote sensing of Environment, 81(1), 149-161.

Estelles, V., Campanelli, M., Utrillas, M. P., Expósito, F., \& Martínez-Lozano, J. A. (2012). Comparison of AERONET and SKYRAD4. 2 inversion products retrieved from a Cimel CE318 sun photometer. Atmospheric Measurement Techniques, 5(3), 569.

Fernald, F. G. (1984). Analysis of atmospheric lidar observations- Some comments. Applied optics, 23(5), 652-653.

Grabner, M., \& Kvicera, V. (2011). The wavelength dependent model of extinction in fog and haze for free space optical communication. Optics express, 19(4), 3379-3386.

Grob, S., Esselborn, M., Weinzierl, B., Wirth, M., Fix, A., \& Petzold, A. (2013). Aerosol classification by airborne high spectral resolution lidar observations. Atmospheric chemistry and physics, 13(5), 2487-2505.

He, Q., Li, C., Geng, F., Zhou, G., Gao, W., Yu, W., Li, Z., \& Du, M. (2016). A parameterization scheme of aerosol vertical distribution for surface-level visibility retrieval from satellite remote sensing. Remote Sensing of Environment, 181, 1-13.

Kim, I. I., McArthur, B., \& Korevaar, E. J. (2001). Comparison of laser beam propagation at 785 nm and 1.55 $\mu m$ in fog and haze for optical wireless communications. In Information Technologies 2000 (pp. 26-37). International Society for Optics and Photonics.

Klett, J. D. (1981). Stable analytical inversion solution for processing lidar returns. Applied Optics, 20(2), 211-220.

Kruse, P. W., McGlauchlin, L. D., \& McQuistan, R. B. (1962). Elements of infrared technology: Generation, transmission and detection. New York: Wiley, 1962.

Liang, S., Zhong, B., \& Fang, H. (2006). Improved estimation of aerosol optical depth from MODIS imagery over land surfaces. Remote Sensing of Environment, 104(4), 416-425.

Liao, S. K., Yong, H. L., Liu, C., Shentu, G. L., Li, D. D., Lin, J., Dai, H., Zhao, S. Q., Li, B., Guan, J. Y., Chen, W., Gong, Y. H., Li, Y., Lin, Z. H., Pan, G. S., Pelc, J. S., Fejer, M. M., Zhang, W. Z., Liu, W. Y., Yin, J., Ren, J. G., Wang, X. B., Zhang, Q., Peng, C. Z., \& Pan. J. W. (2017). Long-distance free-space quantum key distribution in daylight towards inter-satellite communication. Nature Photonics.

Lisenko, S. A., Kugeiko, M. M., \& Khomich, V. V. (2016). Multifrequency lidar sounding of air pollution by particulate matter with separation into respirable fractions. Atmospheric and Oceanic Optics, 29(3), 288-297.

Luo, T., Wang, Z., Ferrare, R. A., Hostetler, C. A., Yuan, R., \& Zhang, D. (2015). Vertically resolved separation of dust and other aerosol types by a new lidar depolarization method. Optics express, 23(11), 14095-14107.

Matzler, C. (2002). MATLAB functions for Mie scattering and absorption, version 2. IAP Res. Rep, 8, 1-24.

Mayor, S. D., Spuler, S. M., Morley, B. M., \& Loew, E. (2007). Polarization lidar at 1.54 $\mu m$ and observations of plumes from aerosol generators. Optical Engineering, 46(9), 096201-096201.

McKendry, I. G., Van der Kamp, D., Strawbridge, K. B., Christen, A., \& Crawford, B. (2009). Simultaneous observations of boundary-layer aerosol layers with CL31 ceilometer and 1064/532 nm lidar. Atmospheric Environment, 43(36), 5847-5852.

Nakajima, T., Tonna, G., Rao, R., Boi, P., Kaufman, Y., \& Holben, B. (1996). Use of sky brightness measurements from ground for remote sensing of particulate polydispersions. Applied Optics, 35(15), 2672-2686.

Shangguan, M., Xia, H., Wang, C., Qiu, J., Shentu, G., Zhang, Q., Dou, X., \& Pan, J. W. (2016). All-fiber upconversion high spectral resolution wind lidar using a Fabry-Perot interferometer. Optics Express, 24(17), 19322-19336.

Shentu, G. L., Pelc, J. S., Wang, X. D., Sun, Q. C., Zheng, M. Y., Fejer, M. M., Zhang, Q., \& Pan, J. W. (2013). Ultralow noise up-conversion detector and spectrometer for the telecom band. Optics express, 21(12), 13986-13991.

Steven, M. D. (1998). The sensitivity of the OSAVI vegetation index to observational parameters. Remote Sensing of Environment, 63(1), 49-60.

USAF, U. (1976). Standard atmosphere, 1976. US Government Printing Office, Washington, DC.

Viezee, W., Uthe, E. E., \& Collis, R. T. H. (1969). Lidar observations of airfield approach conditions: an exploratory study. Journal of Applied Meteorology, 8(2), 274-283.

Weitkamp, C. (Ed.). (2006). Lidar: range-resolved optical remote sensing of the atmosphere (Vol. 102). Springer Science \& Business.

Werner, C., Streicher, J., Leike, I., \& Munkel, C. (2005). Visibility and cloud lidar. In Lidar (pp. 165-186). Springer New York.

Wu, S., Song, X., Liu, B., Dai, G., Liu, J., Zhang, K., Qin, S., Hua, D., Gao, F., \& Liu, L. (2015). Mobile multi-wavelength polarization Raman lidar for water vapor, cloud and aerosol measurement. Optics express, 23(26), 33870-33892.

Xia, H., Dou, X., Sun, D., Shu, Z., Xue, X., Han, Y., Hu, D., Han, Y., \& Cheng, T. (2012). Mid-altitude wind measurements with mobile Rayleigh Doppler lidar incorporating system-level optical frequency control method. Optics express,20(14), 15286-15300.

Xia, H., Shentu, G., Shangguan, M., Xia, X., Jia, X., Wang, C., Zhang, J., Jason, S. P., M. M. Fejer, Zhang, Q., Dou, X., \& Pan, J. (2015). Long-range micro-pulse aerosol lidar at 1.5 μm with an upconversion single-photon detector. Optics letters, 40(7), 1579-1582.

Xia, H., Shangguan, M., Wang, C., Shentu, G., Qiu, J., Zhang, Q., Dou, X., \& Pan, J. W. (2016). Micro-pulse upconversion Doppler lidar for wind and visibility detection in the atmospheric boundary layer. Optics Letters, 41(22), 5218-5221.

Xia, H., Shangguan, M., Shentu, G., Wang, C., Qiu, J., Zheng, M., Xie, X., Dou, X., Zhang, Q., \& Pan, J. W. (2016). Brillouin optical time-domain reflectometry using up-conversion single-photon detector. Optics Communications, 381, 37-42.

Xia, H., Sun, D., Yang, Y., Shen, F., Dong, J., \& Kobayashi, T. (2007). Fabry-Perot interferometer based Mie Doppler lidar for low tropospheric wind observation. Applied optics, 46(29), 7120-7131.

Yin, J., Cao, Y., Li, Y., Liao, S., Zhang, L., Ren, J., Cai, W., Liu, W., Li, B., Dai, H., Li, G., Lu, Q., Gong, Y., Xu, Y., Li, S., Li, F., Yin, Y., 
Jiang, Z., Zhang, X., Wang, N., Chang, X., Zhu, Z., Liu, N., Chen, Y., Lu, C., Shu, R., Peng, C., Wang, J., \& Pan, J. W. (2017). Satellite-based entanglement distribution over 1200 kilometers.Science, 356(6343), 1140-1144.

Yu, C., Shangguan, M., Xia, H., Zhang, J., Dou, X., \& Pan, J. W. (2017). Fully integrated free-running InGaAs/InP single-photon detector for accurate lidar applications. Optics Express, 25(13), 14611-14620.

\end{document}